\documentclass[aps,pre,reprint]{revtex4-1}

\usepackage[utf8]{inputenc}
\usepackage[T1]{fontenc}

\usepackage[main=english,ngerman,latin]{babel}

\usepackage{letltxmacro}

\makeatletter
\LetLtxMacro{\ORIG@selectlanguage}{\selectlanguage}
\DeclareRobustCommand{\selectlanguage}[1]{%
  \@ifundefined{alias@\string#1}
    {\ORIG@selectlanguage{#1}}
    {\begingroup\edef\x{\endgroup
       \noexpand\ORIG@selectlanguage{\@nameuse{alias@#1}}}\x}%
}
\newcommand{\definelanguagealias}[2]{%
  \@namedef{alias@#1}{#2}%
}
\makeatother

\definelanguagealias{en}{english}
\definelanguagealias{EN}{english}
\definelanguagealias{de}{ngerman}
\definelanguagealias{la}{latin}

\usepackage{amsfonts}
\usepackage{amsmath}
\usepackage{color,graphicx}

\usepackage{siunitx}
\DeclareSIUnit\atm{atm}

\begin{document}

\title{Explosive, oscillatory and Leidenfrost boiling at the nanoscale}

\thanks{Published in \emph{Phys Rev E} \textbf{99}, 063110, 27 June 2019; DOI: 10.1103/PhysRevE.99.063110}
\author{Thomas Jollans}
\author{Michel Orrit}
\email{orrit@physics.leidenuniv.nl}
\affiliation{Huygens--Kamerlingh Onnes Laboratory, Leiden University, %
Postbus 9504, 2300\,RA~Leiden, The Netherlands}

\begin{abstract}

We investigate the different boiling regimes around a single continuously
laser-heated \SI{80}{\nm} gold nanoparticle and draw parallels to the classical
picture of boiling. Initially, nanoscale boiling takes the form of transient,
inertia-driven, unsustainable boiling events characteristic of a nanoscale
boiling crisis. At higher heating power, nanoscale boiling is continuous, with
a vapor film being sustained during heating for at least up to \SI{20}{\us}.
Only at high heating powers does a substantial stable vapor nanobubble form.
At intermediate heating powers, unstable boiling sometimes takes the form of
remarkably  stable nanobubble oscillations with frequencies between
\SIlist{40;60}{\MHz}; frequencies that are consistent with the relevant size
scales according to the Rayleigh-Plesset model of bubble oscillation, though
how applicable that model is to plasmonic vapor nanobubbles is not clear.
\end{abstract}

\maketitle

\section{Introduction}

The mechanisms involved in boiling of liquids in contact with a heat source are
of crucial importance when it comes to understanding and optimizing heat transfer,
particularly in applications requiring the removal of high heat flux. In recent
years, there has been particular interest in the effect that the use of
`nanofluids'---fluids containing metal nanoparticles that may attach to
device walls \cite{murshed_review_2011,taylor_pool_2009}---and
nanostructured surfaces have on pool-boiling heat transfer into the fluid
\cite{kim_review_2015}. There are many reports of both nanofluids and nanoscale
surface roughness increasing the critical heat flux that a heating device can support.
It is therefore imperative to gain a deeper understanding of boiling at the
nanoscale, a topic we hope to shed some light on here.

\begin{figure}
\centering
\includegraphics{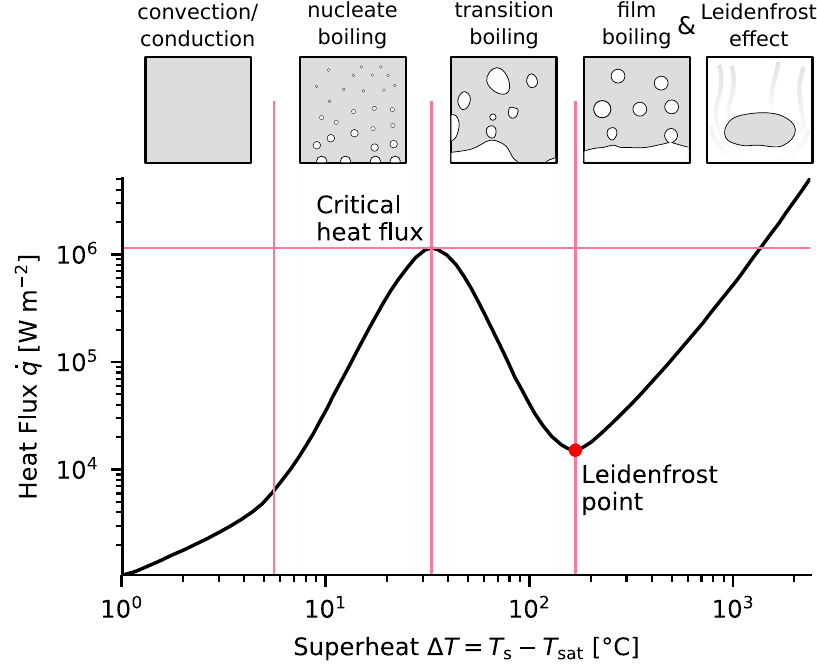}
\caption{
  The well-known traditional boiling curve of macroscopic pools of water at
  atmospheric pressure with sketches of the different boiling regimes.
  Curve data adapted from Ref.\ \cite{cengel_pool_2003}.
}
\label{fig-macro}
\end{figure}

In the canonical model of pool boiling, i.e., boiling of a large `pool' of liquid
through direct contact with a hot surface, boiling is thought to occur in
three primary regimes: in order of increasing relative temperature $\Delta T$,
nucleate boiling, transition boiling, and film boiling (see Fig.~\ref{fig-macro}).
In nucleate boiling, boiling occurs at a myriad microscopic active vapor
generating centers from which small bubbles rise upward (gravity is
significant for common liquids at human size scales), and the resulting total heat flux
from the heating surface to the liquid being boiled is proportional to the number
of active vapor generating centers (bubble nucleation sites) at any given time.
It is well-established that, apart from depending on $\Delta T$, this number depends
on the structure of the heater surface---broadly speaking, rougher surfaces
support more vapor nucleation sites---but how a particular surface geometry
will lead to particular boiling characteristics is not currently understood
\cite{pioro_nucleate_2004,*pioro_nucleate_2004-1, dong_experimental_2014}.

As the temperature and heat flux increase, an ever greater proportion of the surface
will be covered by vapor bubbles. The vapor, with its much lower
thermal conductivity as compared to the corresponding liquid, acts as a thermal
insulator. This leads to the heat flux topping out at a \emph{critical heat flux}
and then falling as the temperature and the vapor coverage of the heater increase.
This phenomenon is known as the \emph{boiling crisis}. The boiling behavior as
the heat flux falls is characterized by large vapor bubbles forming at the heating
surface and rising violently, and is referred to as \emph{transition boiling} or
\emph{unstable film boiling} \cite{kim_review_2015,jakob_warmeubergang_1933,jakob_warmeubergang_1935,cengel_pool_2003}.

If the temperatures are high enough to overcome the thermally insulating effect
of a thin vapor film, boiling can stabilize into so-called \emph{film boiling}.
In the case of small drops of water coming into contact with a larger heating
surface, this leads to drops levitating on a cushion of hot vapor. The heat
flux again increases with temperature; the point of minimum heat flux is known
as the \emph{Leidenfrost point}, and the transition into film boiling is popularly
known as the \emph{Leidenfrost effect}---both named after Johann Gottlob
Leidenfrost, who described the effect in 1756
\cite{leidenfrost_fixitate_1756,*[Translation: ][]{leidenfrost_fixation_1966}}.

While the precise thresholds and dynamics depend on various properties of the
heater, from the material's thermal properties and surface microstructure up to
the macroscopic shape \cite{sher_film_2012}, the broad outline of the behavior
as described above is widely applicable.

In this work, we dive down to the nanoscale using the tools provided by modern
optical microscopy and study submicrosecond boiling dynamics at a single
artificial nucleation site in the form of a laser-heated gold nanoparticle
(AuNP); AuNPs are frequently used to optically generate vapor micro- and
nanobubbles \cite{hou_explosive_2015,baffou_super-heating_2014,%
siems_thermodynamics_2011,boulais_plasma_2012,hashimoto_studies_2012,%
lombard_kinetics_2014,lombard_nanobubbles_2015,setoura_stationary_2017,%
li_oscillate_2017}. We will find striking parallels to the progression of a
macroscopic system through the boiling crisis, in which the entire heater
surface dries out when a thermally insulating vapor film forms, and the heat
flux plummets.

\section{Method}\label{sec:method}

This work follows on from our previously published results
\cite{hou_explosive_2015}, in which we described an unstable, explosive
nanoscale boiling regime arising under continuous heating, near the threshold
heating power for boiling. Using mostly the same technique, here we investigate
in detail the various dynamics arising from a much broader range of different
heating powers, with an emphasis on exploring the parameter space beyond the
threshold.

Gold nanospheres with a diameter of \SI{80}{\nm} (from NanoPartz) are
immobilized on a cover glass at very low surface coverage by spin-coating. The 
nanoparticles are submerged in a large reservoir of \emph{n}-pentane and
investigated optically in a photothermal--confocal microscope described in
previous work \cite{hou_explosive_2015,gaiduk_detection_2010}: two
continuous-wave laser beams, a heating beam and a probe beam, are carefully
overlapped and tightly focused on the same nanoparticle. The heating beam is
partly absorbed by the sample; the deposited energy and associated temperature
increase lead to localized changes in the sample, e.g., in density. These changes
affect how much the probe beam is scattered.

In photothermal microscopy, these small heating-induced changes can be used to
measure a nano-object's absorption cross section
\cite{jollans_nonfluorescent_2019}. In this work, we focus on the dynamics of
the response, specifically in the case of boiling, instead.


\emph{n}-Pentane was chosen as a medium, as in our previous work, due to its
boiling point under ambient conditions (viz.\ ca.\ \SI{36}{\celsius}) being
close to room temperature; the intention of this choice was to reduce the
necessary heating powers and the impact of heating-related damage to the AuNPs.

\begin{figure}
\centering
\includegraphics{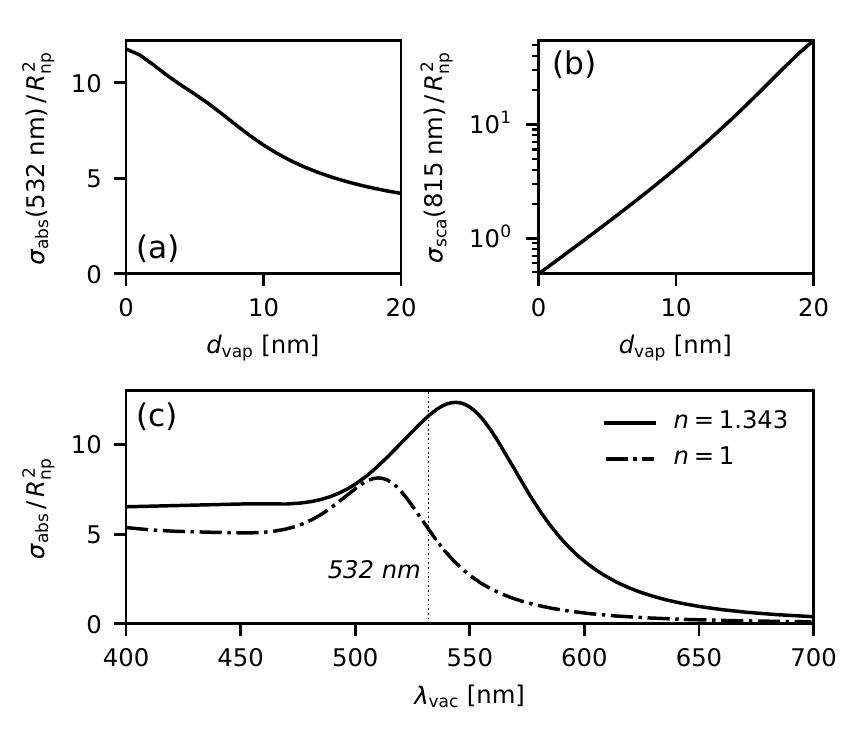}
\caption{Layered-sphere Mie theory calculation \cite{pena_scattering_2009} with
an \SI{80}{\nm} diameter gold core and a vacuum shell (bubble, $n = 1$) of varying
thickness $d_\mathrm{vap}$ in a medium with refractive index $n = 1.33$.
{(a)} Absorption of $\lambda_\mathrm{vac} = \SI{532}{\nm}$ and
{(b)} scattering of $\lambda_\mathrm{vac} = \SI{815}{\nm}$, both shown
as a function of bubble thickness. {(c)} Absorption spectra of
\SI{80}{\nm} AuNPs, when in $n$-pentane as compared to vacuum. See also:
Appendix \ref{app:optics}.
}
\label{fig-mie}
\end{figure}

A single gold nanoparticle---identified through photothermal
contrast---is heated using a focused near-resonant [\SI{532}{\nm}; cf.\ %
Fig.\ \ref{fig-mie}(c)] laser, the intensity of which is controlled using an
acousto-optic modulator (AOM) and monitored using a fast photodiode. The
nanoparticle is monitored in real time through the back-scattering of an
off-resonant probe laser (\SI{815}{\nm}), measured using another fast
photodiode. Our real-time single-nanoparticle optical measurements allow us to
noninvasively follow the dynamics of boiling and vapor nanobubble formation
around the AuNP with a high time resolution, limited by the \SI{80}{\mega\hertz}
cut-off frequency (\SI{-3}{\deci\bel}) of our detector.

It is important to note at this point that a given heating laser intensity
uniquely determines neither the absorbed heating power nor the temperature of
the AuNP\@. Rather, the absorption cross section $\sigma_\mathrm{abs}$ of the AuNP, and with
it the absorbed power, depends strongly on the environment: The localized surface
plasmon resonance of the AuNP depends on the refractive index of the environment,
i.e.,\ on whether the AuNP is surrounded by liquid ($n \approx 1.33$) or vapor
($n \simeq 1$).

Modeling the  nanoparticle surrounded by a vapor layer as a multilayered sphere
with a gold core and a shell with a refractive index of $n = 1$ and with a
certain thickness $d_\mathrm{vap}$, a layered-sphere Mie theory calculation
\cite{pena_scattering_2009} can give us an idea of how
$\sigma_\mathrm{abs}$ changes upon nanobubble growth: Fig.~\ref{fig-mie}(a)
shows $\sigma_\mathrm{abs}$ dropping by \SI{10}{\percent} with only a
\SI{3}{\nm} thick bubble, and by half with \SI{13}{\nm} thickness. This leads to
negative feedback as a vapor shell grows due to its optical properties. This
complements the negative feedback due to the shell's thermal properties which
is known from the classical Leidenfrost effect.



At the same time, the dependence of our read-out, the back-scattering at
\SI{815}{\nm}, on the bubble size is not trivial. As Fig.~\ref{fig-mie}(b) shows,
for sufficiently thin vapor nanobubbles, we can expect the scattering to grow
steeply with the bubble thickness.

The AuNP is initially subjected to a heating laser intensity such that it is near
the boiling threshold, but below it; then, periodically, the AOM is switched to
provide a pulse of some microseconds at a higher illumination intensity (duty
cycle: \SI{1}{\percent}).

The baseline intensity is chosen heuristically on a nanoparticle-by-nanoparticle
basis by testing increasing baselines with a fixed additional on-pulse intensity
until the measured scattering shows a significant change. The chosen baseline
intensity is between \SI{80}{\micro\watt} and \SI{140}{\micro\watt} as measured
in the back focal plane. The pulse length, pulse height, and baseline can then
be changed at will between timetrace acquisitions; the length of a full
timetrace was \SI{10}{\ms}, of which the intensity is ‘high’ for \SI{100}{\us}.
Between subsequent acquisitions, a few seconds of dead time pass while data is
stored and conditions are, as the case may be, changed.

\section{Results}

\subsection{Nanoscale boiling regimes}

Depending on the heating power during the heating pulse, boiling around the
nanoparticles was observed to
follow four distinct patterns as shown in figure~\ref{fig-regimes}, before
irreversible damage sets in at higher powers:
\begin{figure}
  \centering
  \includegraphics{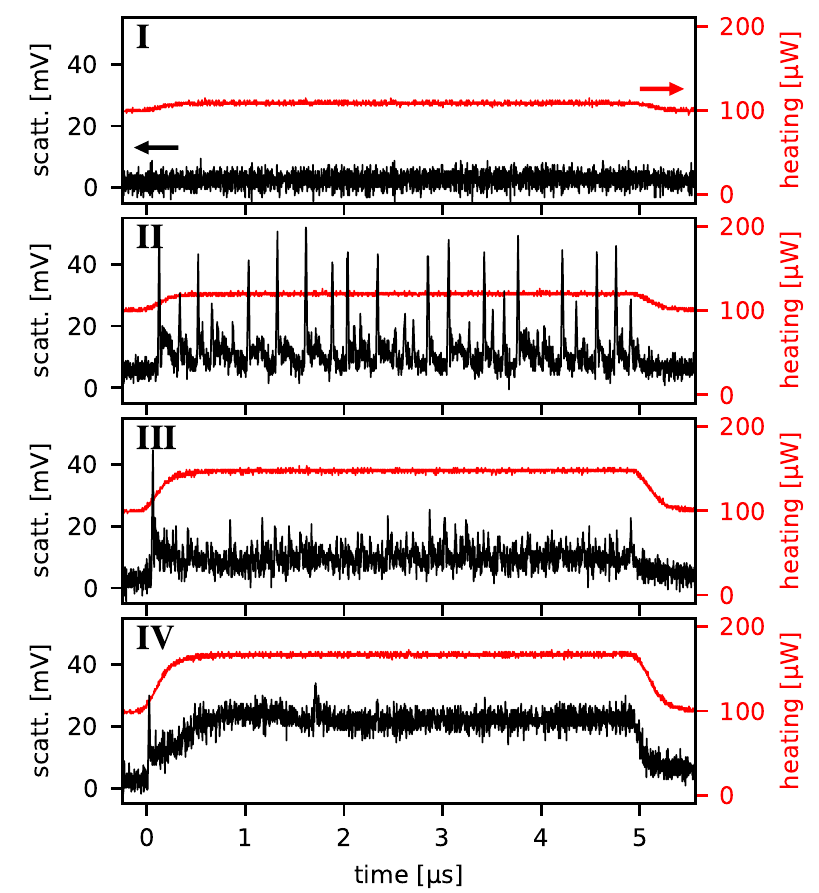}
  \caption{
    Single-shot scattering timetraces of the same \SI{80}{\nm} Au sphere.
    Black: scattering in the detector's units; red: heating intensity in the
    back focal plane. Four principal regimes of heated AuNP behavior are observed:
    (I)~below boiling threshold; small thermal lensing effect.
    (II)~near boiling threshold; repeated short-lived vapor nanobubble formation.
    (III)~above boiling threshold; unstable vapor nanobubble
    for the duration of heating.
    (IV)~far above threshold; relatively stable nanobubble, probable damage to AuNP\@.
    See also: figure~\ref{fig-regime-map}
  }
  \label{fig-regimes}
\end{figure}


I.---At sufficiently low power, the effect of heating the AuNP is limited
to a small thermal lensing effect that cannot easily be directly identified from
the timetrace as shown. There is no indication of boiling.

II.---Above a certain threshold, $\sim\SI{120}{\micro\watt}$ in the back focal
plane with some variation from particle to particle, as previously demonstrated
\cite{hou_explosive_2015}, strong $\sim\SI{1e-8}{\second}$ spikes with similar
peak values start appearing at random intervals. These can be explained by
rapid inertially driven expansion of a vapor nanobubble around the AuNP; in
the presence of this vapor shell, the hot AuNP experiences a boiling crisis and
the nanobubble collapses immediately. While this behavior is reminiscent of
intermittent film boiling, averaged over many disperse vapor generating centers
it should appear as nucleate boiling from a distance.

III.---As the power is increased further, beyond around \SI{150}{\micro\watt},
all distinct explosive spikes but the first disappear. In their place, the
initial explosion is followed by still highly dynamic behavior, notably with a
much smaller amplitude. It appears that while in the previous case, the AuNP
returned to the same state after the boiling events (i.e., no vapor), now, it
does not; after the initial expansion, the nanobubble does not appear to fully
collapse, but rather to reduce to a sustainable if not particularly stable size.
We can think of this as a transition boiling regime.

IV.---At extremely high powers, the signal steadily grows [as expected for
a growing vapor bubble; cf.\ Fig.\ \ref{fig-mie}(b)], after the initial
spike, to a stable level that is maintained until heating ends. The signal
overall appears calmer than in the previous cases. In analogy with the
Leidenfrost effect known macroscopically, it appears that a nanometer-scale
vapor film around the heated AuNP is stabilized only at these substantial
heating powers.

Note that in the explosive regime (II), all events are clearly separated from
one another, and all have approximately the same maximum value. In particular,
no double-peaks have been observed. This indicates that each nanoparticle
hosts only a single vapor generating center.

In sustained boiling regimes (III, IV), nanobubble behavior qualitatively stays
the same from after the initial expansion until the heating intensity is reduced,
for at least up to \SI{20}{\us}: long-lived vapor nanobubbles do not appear to
spontaneously collapse without a change in externally imposed conditions.

However, extended or repeated irradiation at powers sufficient for boiling will
cause irreversible damage to the nanoparticle: in particular, after minutes of
irradiation at high power, lower powers no longer show the familiar
explosive nucleate boiling events. We cannot tell in what way the particles
change during long experiments. They might be melting, which could involve
changing surface structure and/or contact area with the substrate, they might
be fragmenting \cite{setoura_cw-laser-induced_2014}, or they may be sinking
into the substrate \cite{hashimoto_gold_2009}. Any one of these
possibilities would change the optical and thermal properties in
hard-to-predict ways.

%
%
\begin{figure}
  \centering
  \includegraphics[width=8cm]{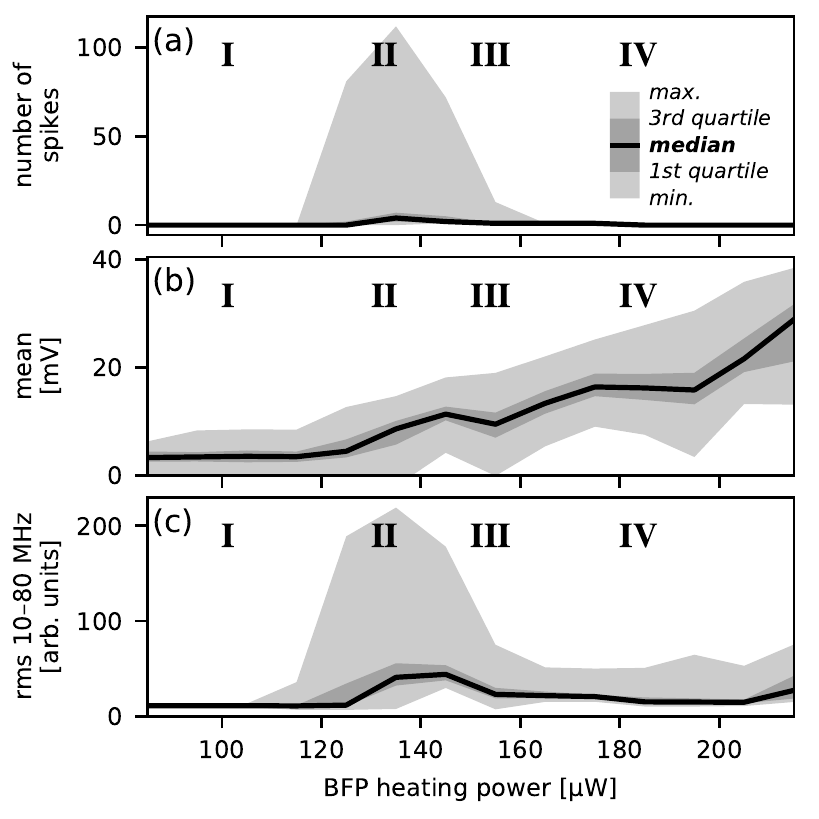}
  \caption{
    Dependence of vapor nanobubble behavior on laser power, based on 8050
    single-shot time traces measured at one nanoparticle, like those shown in
    figure~\ref{fig-regimes}. {(a)} number of
    spikes (extrema $\mathrm d \tilde u \big/ \mathrm d t = 0$ where
    $\mathrm d^2 \tilde u \big/ \mathrm dt^2 < 0$ is below a heuristically chosen
    threshold, $\tilde u(t)$ being the signal smoothed with a \SI{30}{\nano\second}
    Hann filter) in the signal. {(b)} mean scattering signal
    during heating. {(c)} RMS power of the frequency components between
    \SIlist{10;80}{\mega\Hz}.
  }
  \label{fig-regime-map}
\end{figure}

Figure~\ref{fig-regime-map} shows how the behavior varies with heating power;
in particular, the explosive regime (II) clearly occurs only in a limited
temperature range. Note also that the mean response does not have an equivalent
to the nucleate boiling peak in the classical curve from figure~\ref{fig-macro}.

\subsection{Stable vapor nanobubble oscillations}

\begin{figure*}[bt]
  \makebox[\textwidth][c]{
    \includegraphics{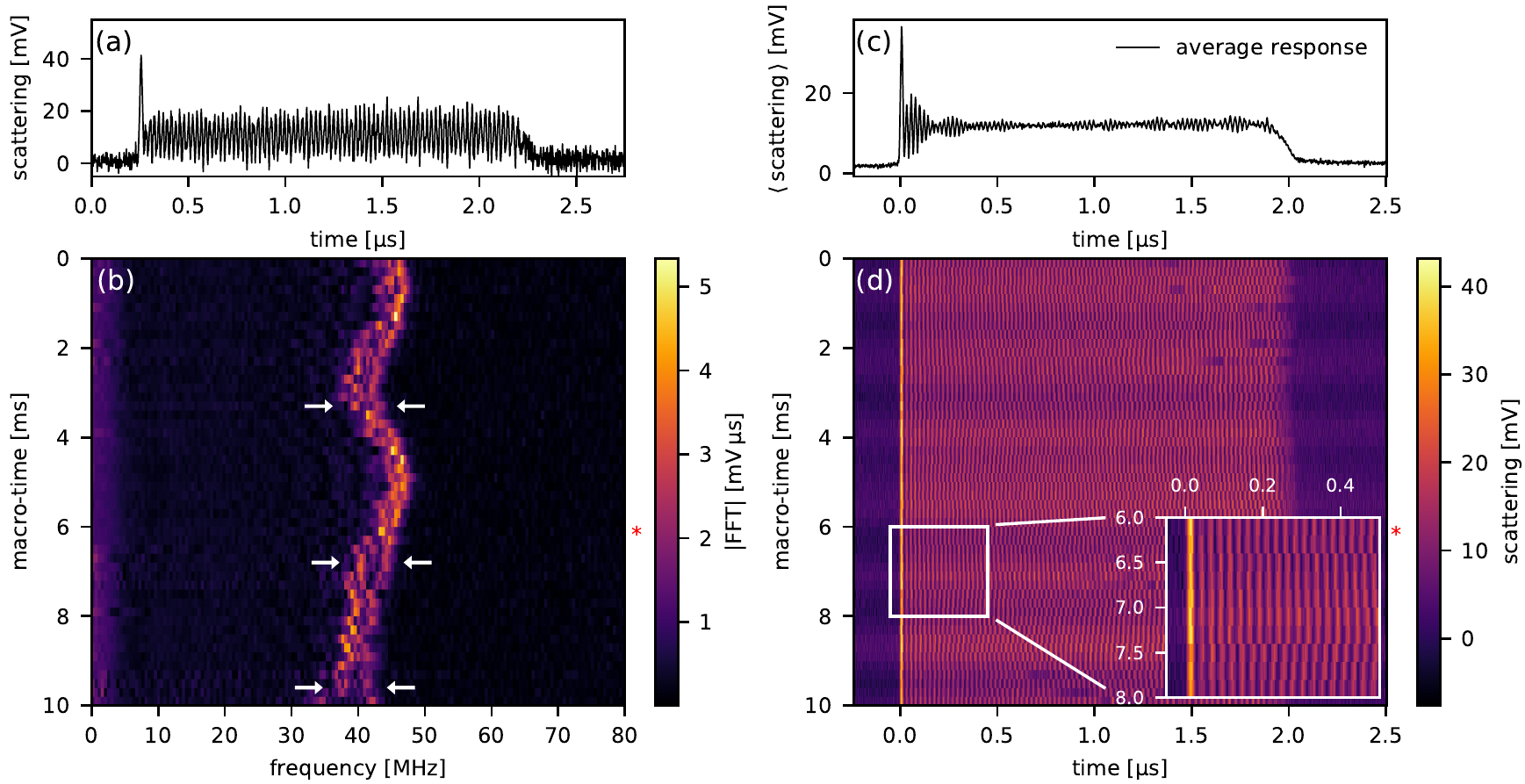}
  }
  \caption{
    Nanobubble oscillations;
      (a) Single-shot nanobubble timetrace with clearly visible
        oscillations.
      (b) Fourier transforms of 50 subsequent such timetraces of the
        same AuNP subject to an unchanging heating profile. White arrows
        highlight some points with particularly clear frequency splitting.
      (c) Average of the timetraces in panel (d), synchronized on the initial
        rising edge. 
      (d) Fifty timetraces used in panels (b) and (c). In panels (b) and (d), the lines
        marked with an asterisk correspond to panel (a).
  }
  \label{fig-osci}
\end{figure*}

For some nanoparticles, heating powers characterized by unstable boiling were
found to produce not a randomly growing and collapsing vapor nanobubble, but a
stable and surprisingly pure oscillation with frequencies around
\SI{50 \pm 10}{\mega\Hz} (see Fig.~\ref{fig-osci}).

To crudely model the oscillations, we shall employ the Rayleigh-Plesset model
\cite{lauterborn_physics_2010} for spherical gas bubble oscillations, which has
been shown theoretically to be remarkably effective for describing the kinetics
of the initial expansion and collapse of a plasmonic vapor nanobubble
\cite{lombard_kinetics_2014,lombard_nanobubbles_2015}:
\begin{align}
\varrho R \ddot R + \frac{3}{2}\varrho \dot R ^2 ~=~&
  p_\mathrm{g} \left(\frac{R_0}{R}\right)^{3\kappa}
  - p_\mathrm a - \frac{\mathrm{2\gamma}}{R}
  - \frac{3\mu}{R}\dot R, \label{eq:RP}\\
\mathrm{with}~~~p_\mathrm{g}~=~& \frac{2\gamma}{R_0} + p_\mathrm a,
\end{align}
where $R$ is bubble radius, $R_0$ is the equilibrium radius, $\kappa = c_p/c_V$
is the polytropic exponent, $\gamma$ is the surface tension, $\mu$ the dynamic
viscosity, $\varrho$ the liquid density, $p_\mathrm a$ the ambient (static)
pressure, and $p_\mathrm g$ the gas bubble pressure at rest.

Compared to the full Rayleigh-Plesset equation as given by
\citet{lauterborn_physics_2010}, we are assuming no external ultrasound field
and make no distinction between vapor and gas pressure. The latter approximation
is an adiabatic approximation---it forbids mass transfer between the bubble
and the liquid at the relevant timescales. If the bubble were to be understood
classically as vapor, meaning vapor molecules can freely condense into the
surrounding liquid, then there would be no restorative force due to compression
when the size of the bubble is reduced; there could be no oscillation. 
Oscillations are, however, clearly observed. Hence, we proceed assuming full
conservation of mass for the material inside the bubble, i.e., we
treat the vapor nanobubble as a classical gas bubble. (N.B., the applicability of
the model is further discussed in Appendix \ref{app:rp}.)

Expanding eq.~\eqref{eq:RP} for small perturbations $R = R_0 (1 + \varepsilon)$
from the equilibrium to first order in $\varepsilon$, we can reduce the
Rayleigh-Plesset model to a damped harmonic oscillator,
\begin{align}
\frac{\mathrm d^2 \varepsilon}{\mathrm d t^2} & + 2\zeta\omega_0 \frac{\mathrm d \varepsilon}{\mathrm d t} + \omega_0^2 \varepsilon = 0,\\
\mathrm{where}~
\omega_0~=~& \frac{1}{R_0 \sqrt \varrho} \sqrt{
    3\kappa p_\mathrm a + \frac{2\gamma}{R_0}\big(3 \kappa - 1\big)
    } \label{eq:omega0}\\
\mathrm{and}~
\zeta~=~& \frac{3\mu}{2\varrho R_0^{\,2} \; \omega_0 } . \label{eq:zeta}
\end{align}

\begin{figure}\centering
\includegraphics{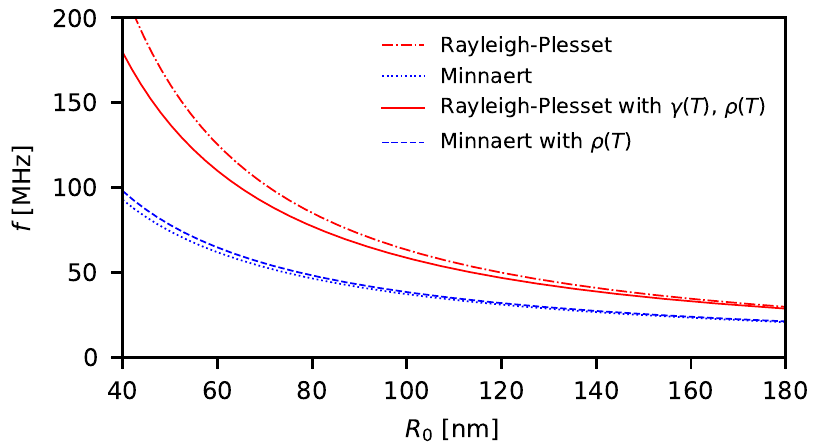}
\caption{Resonance frequencies for vapor nanobubbles in pentane, according to
the Rayleigh-Plesset and Minnaert models, with and without considering the
temperature dependence of the surface tension $\gamma$ and the density $\varrho$.}
\label{fig:RP}
\end{figure}

This allows us to calculate the resonance frequency
$f = \omega_0 \sqrt{1 - \zeta^2} / 2 \pi$, shown in Fig.~\ref{fig:RP}, using the
well-known material properties of pentane \cite{noauthor_pentane_2018} at the
saturation point \cite{vesovic_pentane_2011} at $p_\mathrm{a} = \SI{1}{\atm}$:
$f = \SI{40}{\mega\hertz}$ corresponds to $R_0 = \SI{142}{\nano\meter}$. The
influence of viscous damping has a negligible impact on the resonance frequency
as $\zeta \sim 0.1$ is small; in the real system, damping seems to be
counteracted by the driving force from heating, leading to stable
self-oscillation.

We can take into account the temperature- and pressure-dependence of the density
$\varrho(T,p_\mathrm{g})$ and surface tension $\gamma(T,p_\mathrm{g})$
by using the values for saturated liquid at
$p_\mathrm{g} = p_\mathrm{a} + 2\gamma/R_0$ and the saturation temperature
$T = T_\mathrm{sat}(p_\mathrm{g})$. If we do this, the results are only slightly
changed: then, $f = \SI{40}{\mega\hertz}$ corresponds to
$R_0 = \SI{136}{\nano\meter}$.

It is interesting to note that the two terms under the square root in eq.\ %
\eqref{eq:omega0}, $3\kappa p_\mathrm a = \SI{0.33}{\mega\pascal}$ and
$2\gamma(3\kappa-1)/\SI{136}{\nm} = \SI{0.36}{\mega\pascal}$, are of the same order
of magnitude. The resulting frequencies, then, are not dramatically different from
those predicted by the much simpler Minnaert model. In 1933, Marcel
Minnaert proposed a simple model to explain the `musical', i.e., audible,
oscillations of spherical air bubbles in a stream of water (e.g. from a tap)\
\cite{minnaert_musical_1933}, which does not consider surface tension or
viscous drag, only the compressibility of the gas:
\begin{equation}
\omega_\mathrm{Minnaert} = \frac{1}{R_0} \sqrt{\frac{3 \kappa p_\mathrm{a}}{\varrho}}
\end{equation}

In any case, all oscillation frequencies observed correspond to radii larger
than the radius of the AuNP (viz. $R_\mathrm{np} = \SI{40}{\nano\meter}$),
up to approximately the size of the near diffraction-limited
focus of the heating laser (viz. FWHM $w_\mathrm\perp=\SI{0.2}{\um},~
w_\mathrm{\parallel}=\SI{0.6}{\um}$). Direct
measurement of bubble size is not possible with the present
technique, but these estimates are in agreement with previous measurements
of vapor nanobubble sizes \footnote{See also Sec.\
11 of the Supplementary Information of Ref. \cite{hou_explosive_2015}.}.

\begin{figure}
\centering
\includegraphics{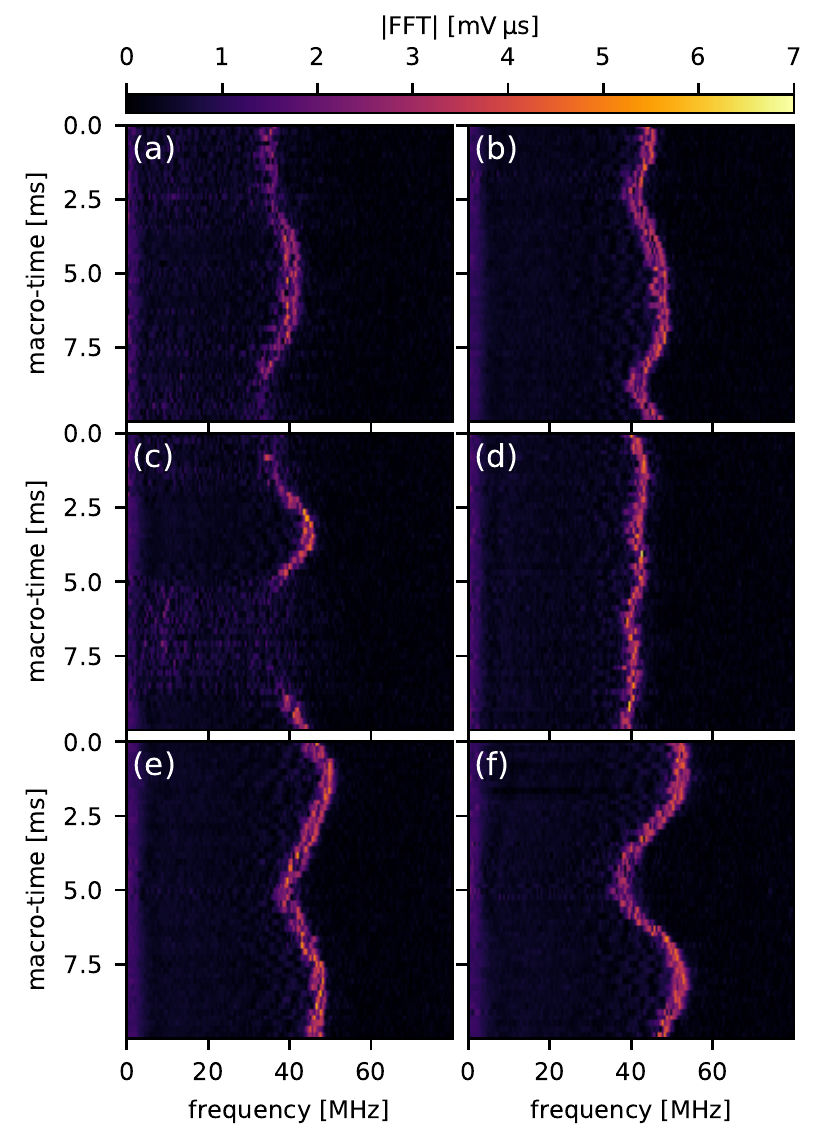}
\caption{Example series of oscillating-bubble event Fourier transforms, all for
the same nanoparticle, in the same form as Fig.~\ref{fig-osci}(b).}
\label{fig-osci-ffts}
\end{figure}

The factors contributing to the oscillations are evidently not random: as
Figs.\ \ref{fig-osci}(b)--\ref{fig-osci}(d) show, the frequencies are strongly correlated from one
event to the next; the resonance frequency appears to drift back and forth over
time at audio frequencies, perhaps as a response to acoustic noise or small
vibrations in the microscope. As the other examples in
Fig.~\ref{fig-osci-ffts} demonstrate, both the rate and periodicity of the
frequency drift vary from measurement to measurement.

\begin{figure}
\centering
\includegraphics{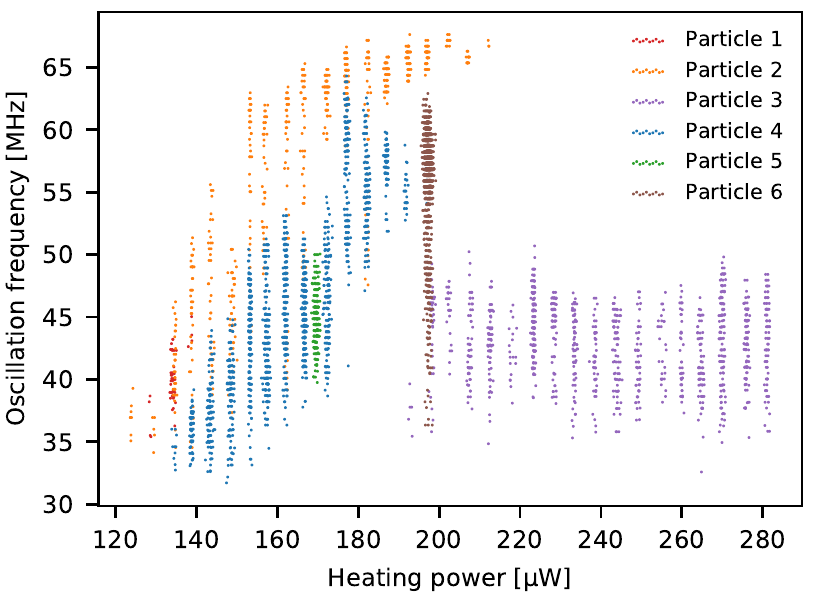}
\caption{Change of the apparent oscillation frequency with heating power for
different AuNPs. This figure shows the location of peaks in the Fast Fourier
Transforms of individual time traces such as those shown in
Fig.\ \ref{fig-osci}(b). Only measurements showing good oscillations are shown:
those with $Q > 10$, where the quality factor
$Q = f_\mathrm{max}/\mathrm{FWHM}$ is calculated from the FFT.
``Particle 4'' refers to the set of measurements used in figures \ref{fig-osci},
\ref{fig-osci-ffts} and \ref{fig:osci-ffts-harmonics}.
\label{fig:osci-pwrs}}
\end{figure}

Additionally, as shown in figure \ref{fig:osci-pwrs}, oscillation frequencies
vary from particle to particle, as well as from moment to moment under constant
experimental conditions. For some, but not all, nanoparticles, the mean
oscillation frequency appears to increase with heating power. However, as all
the measurements were taken sequentially from low to high heating power, the
changes in frequency may be due to ageing of the nanoparticle rather than
due to any heating-dependent effect.

The Fourier transforms [Fig.\ \ref{fig-osci}(b)] of many events show two
frequencies split by a few \si{\mega\hertz}. In real space, this corresponds to
a beat note which is visible faintly in a single timetrace [e.g.\ Fig.\ %
\ref{fig-osci}(a)] and visibly very clearly in the mean of a series of events,
shown in Fig.\ \ref{fig-osci}(c).  Note that the timetraces in
Figs.\ \ref{fig-osci}(c) and \ref{fig-osci}(d) are synchronized on the initial
rising edge of the response, not the heating pulse, in order to eliminate the
possible effect of jitter in initial explosion. The fact that the mean in
Fig.\ \ref{fig-osci}(c) clearly shows the first few periods of the oscillation
demonstrates that the oscillations are very consistently in phase from one
event to the next.

The oscillations appear to only be stable between roughly
\SIlist{40;60}{\mega\hertz}. From time to time (e.g.\ in
Fig.~\ref{fig-osci-ffts}c), it looks like the resonance frequency drifts below
\SI{40}{\mega\hertz} and the oscillation collapses, before resuming later.

Oscillations manifested themselves only around some of the AuNPs tested,
but were remarkably robust against changes in power---leaving aside
aforementioned heating-induced damage to the AuNP\@. In contemplating why these
oscillations might only appear some of the time, we find ourselves confronted
with the question of how the oscillations are possible at all:
Existing models and reports of bubble oscillation \cite{lauterborn_physics_2010},
including at very small scales \cite{macdonald_oscillating_2003,li_oscillate_2017},
describe the oscillation of gas, rather than pure vapor bubbles.

Here, however, any sign of the bubble, or indeed its oscillation, disappear
as soon as the heating ceases. This makes the notion of a permanent gas core in
the oscillating bubbles seem unlikely. As a AuNP that supports oscillating vapor
nanobubbles does so consistently---brief interruptions as seen in
Fig.~\ref{fig-osci-ffts}c notwithstanding---we suspect that whether any
particular AuNP supports these oscillations is linked to unknown structural or
geometric properties, e.g., their exact volume, the size and structure of the
facets on their surfaces, or the contact area with the glass slide.

\section{Conclusion}

By scaling down a heating element to the nanoscale, we have, simultaneously,
scaled down the classical boiling regimes, from nucleate boiling to partial and
full film boiling. 

The nucleate boiling regime is stunted; rapid inertially driven expansion of
insulating vapor bubbles leads the system 
into a boiling crisis, where the absorbed power is insufficient to drive
continued boiling in the presence of the newly formed vapor layer.

At higher incident powers, a boiling regime reminiscent of unstable film
boiling can be sustained. Nanobubble oscillations \emph{can} then be driven by
the nanoheater, but for the most part, unstable boiling at the nanoscale is
characterized by random fluctuations. When the laser intensity is sufficient
for the AuNP to absorb and transduce a critical heat flux, even while
surrounded by a thin vapor shell, vapor bubble formation stabilizes itself,
leading to a nanoscale Leidenfrost effect.

Vapor nanobubble oscillations, when they occur, are remarkably consistent
with the canonical model, the Rayleigh-Plesset equation, for oscillating gas
bubbles of a similar size in the same environment. It would appear that, under
certain conditions, vapor bubble dynamics are faster than vapor--liquid
equilibration.

The transition from a highly unstable or explosive boiling regime to a stable
one may have ramifications for potential applications of highly-heated
nanoparticles. Mechanical stresses caused by bubble formation around gold
nanoparticles, thought to be relevant in the context of plasmonic photothermal
therapy \cite{zharov_microbubbles-overlapping_2005,huang_plasmonic_2008,%
lapotko_plasmonic_2011}, may well be greater in an unstable boiling
scenario compared to a stable one. The intuitive maxim that more laser power
leads to more damage may, under these circumstances, not apply---just as
the relation between heat flux and temperature in macroscopic systems has long
been known to be nontrivial.

\begin{acknowledgments}

The authors would like to thank Julien Lombard and Samy Merabia for useful
discussions. The authors acknowledge funding from the Zwaartekracht program
``Frontiers of Nanoscience (NanoFront)'' of the Netherlands Organisation for
Scientific Research (NWO).

\end{acknowledgments}

\appendix

\section{Approximations: Optics}
\label{app:optics}

The Mie-theoretical treatment of the optical properties of a multilayered
sphere is exact for perfect sphere in an isotropic environment, for an incident
plane-wave field.

On the first point: The nanoparticle is very nearly spherical. A transient
vapor nanobubble is \emph{presumably} quite spherical in order to minimize
surface area. However, the environment is not isotropic in our case; the AuNP
is located on a glass surface (and illuminated from below, through the glass).

On the second point: the approximately Gaussian beams are tightly focused to
near the Abbe diffraction limit. For the nanoparticle itself, the finite size
of the beam is negligible. When a vapor bubble approaches the size of the focus,
however, the finite size of the beam will have a greater impact, further
complicating the optical problem.

The straightforward treatment of the cross sections $\sigma_\mathrm{abs},~%
\sigma_\mathrm{sca}$ does not take into account possible (de)focusing of the
beams by a nanobubble.

\section{Approximations: Rayleigh-Plesset model}
\label{app:rp}

The Rayleigh-Plesset equation assumes the spherical bubble is composed of an
ideal gas and that there is no exchange of material between the bubble and the
liquid (no evaporation, condensation, dissolution, mixing, etc.). Evaporation
and condensation can be included by including a vapor pressure term in the
static pressure.

By not including a vapor pressure term, we are requiring full conservation of
mass in the bubble. Some degree of conservation of mass is required to give rise
to a restorative force and hence oscillations, as indicated above.

\citet{lauterborn_physics_2010} include partial exchange of mass in their
Rayleigh-Plesset equation by introducing a vapor pressure $p_\mathrm v$ that
does not contribute to the pressure on the bubble wall. This is accomplished by
replacing all occurrences of $p_\mathrm a$ in our Eqs.\ (1) and (2) with
$p_\mathrm a - p_\mathrm v$. This would, all said and done, reduce the effective
ambient pressure, thereby lowering the resonance frequency for any given radius.
In this light, the radii derived above might be understood as rough
upper bounds.

Besides the possibility of a vapor bubble with partial conservation of (vapor)
mass, one might consider a mixed vapor-gas bubble. In this case, the nonvapor
gas would provide the restorative force generating the mechanical resonance.
However, as we point out above, we do not believe this explanation is
compatible with the fact that all signs of the bubble, including the
oscillations, disappear when heating ends.

It further does not take into account damping through sound radiation, any
temperature dependence, or deviation from spherical symmetry. No solid gold
object in the center is accounted for in the model, either. However, for a
nanoparticle with $R_\mathrm{np} = \SI{40}{\nm}$ and a bubble with $R_\mathrm b
\approx \SI{120}{\nm}$, the volume of the nanoparticle is less than
\SI{4}{\percent} of the bubble volume. The presence of the AuNP can therefore
be neglected.


With regard to the question of why the vapor in the bubble is compressible
\emph{at all}, i.e.,\ why vapor molecules do not appear to simply condense
into the liquid when the bubble contracts, we can estimate the mean free path
of a pentane molecule:
\[
\ell = \frac{k_\mathrm B T}{\sqrt2 \pi d^2 p},
\]
where $d$ is the molecular diameter and $p$ is the pressure. Taking
$d=\SI{0.43}{\nm}$ \cite{funke_separations_1997}, $T = T_\mathrm{sat}$ and
$p = p_\mathrm{sat} = \SI{1}{atm} + 2\gamma/R$ ($\gamma$ being the
surface tension at saturation \cite{vesovic_pentane_2011} and
$R = \SI{120}{\nm}$ being the nanobubble radius), we get a mean free path
of $\ell = \SI{18.4}{\nano\meter}$.

This is smaller than the bubble thickness, meaning that the dynamics of the
molecules deep in the vapor layer are not affected by the presence of the
vapor--liquid interface and it is plausible that these molecules may contribute
to a restorative pressure just as foreign gas molecules would. This reasoning
is not valid, of course, for the outer quarter or so of the bubble. 

More broadly, this mean free path gives us a Knudsen number of
order $\mathrm{Kn} \sim 10^{-1}$, confirming that a continuum hydrodynamic model
like the Rayleigh-Plesset model can be applied to the bubble. For further
confirmation that the continuum approximation applies, we can estimate that in
an $R = \SI{100}{\nm}$ sphere of a gas with \SI{22.4}{\liter\per\mole}, we expect
some $\sim 10^5$ molecules.

More complex models can include more accurate equations of state (such as a
van der Waals gas law), sound radiation, and other terms
\cite{lauterborn_physics_2010}. \citet{lombard_nanobubbles_2015} have described
a detailed model of the heat transfer problem, but even that cannot account for
the optical feedback expected in the case of continuous-wave heating, touched
upon in Sec.\ \ref{sec:method}.


\begin{figure*}
\centering
\includegraphics{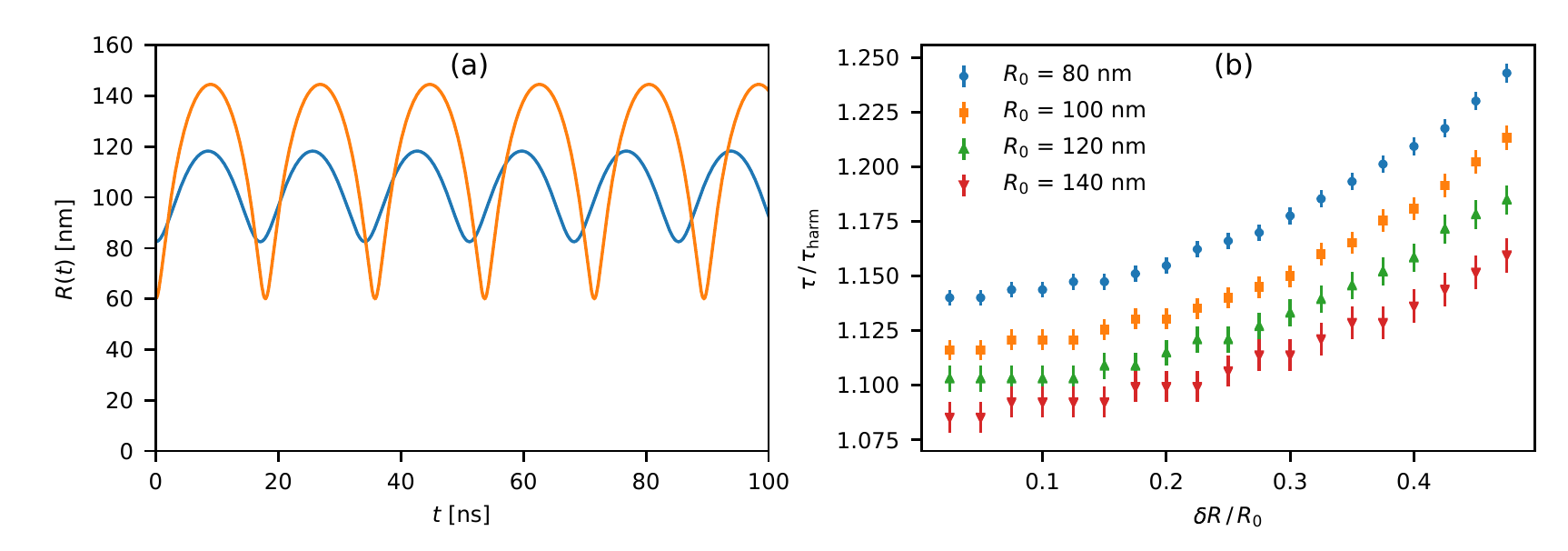}
\caption{
    Anharmonicity of the Rayleigh-Plesset equation for a freely oscillating
    bubble, with the viscosity set to zero. Calculated by direct numerical (Runge--Kutta) integration with $\Delta t = \SI{10}{\pico\second}$ using the
    parameters for saturated pentane. {(a)} Example timetraces of a
    bubble with $R_0 = \SI{120}{\nm}$ oscillating at two different amplitudes.
    The bottom of the curve, especially at larger amplitude, is noticeably more
    `pointed', and the oscillation period is clearly different.
    {(b)} Calculated oscillation periods $\tau$ with different
    equilibrium radii $R_0$ and oscillation amplitudes $\delta R$, relative
    to the corresponding period $\tau_\mathrm{harm}$ in the harmonic approximation.
\label{fig:rp-anharmonic}
}
\end{figure*}

Further, our harmonic approximation drops all higher-order terms. The full equation
predicts some anharmonicity at larger deviations, as shown in fig.\ %
\ref{fig:rp-anharmonic}. The error in the oscillation frequency predicted
using the harmonic approximation at larger amplitudes is presumably small
compared to the unclear effect of the driving force (due to heating) and of
condensation and evaporation.

\begin{figure*}
\centering

\includegraphics{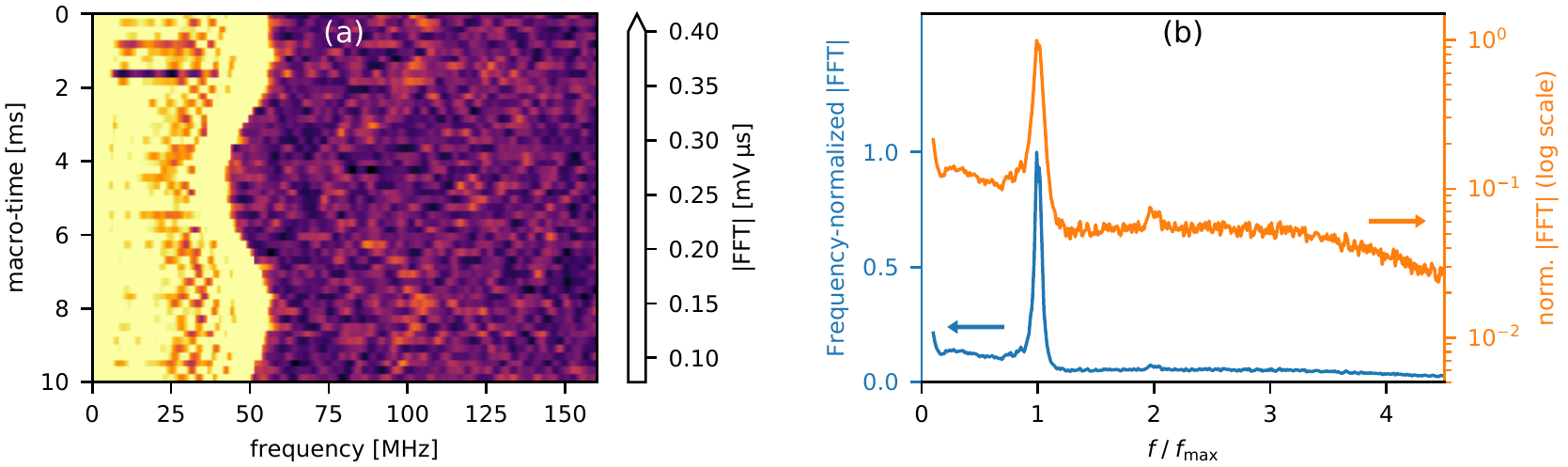}

\caption{{(a)} Series of oscillating-bubble event Fourier transforms\
[same data as Fig.\ \ref{fig-osci-ffts}(f), smoothed with a
\SI{6}{\mega\hertz} Hann filter] showing, faintly, the second harmonic.
{(b)} Mean of the Fourier transforms shown,
calculated after rescaling the frequency axis of each to put the maximum at
unity, showing clearly the second harmonic. \\
Note that the upper cut-off frequency of our detector is
\SI{80}{\mega\hertz}, which reduces the apparent prominence of the second
harmonic, and completely obscures higher harmonics.
\label{fig:osci-ffts-harmonics}}
\end{figure*}

A slight anharmonicity is visible in the measured oscillations: in the series
of Fourier transforms shown in fig.\ \ref{fig:osci-ffts-harmonics}, for
example, the second harmonic of the oscillation is just discernible when the
Fourier transform is smoothed, or in the mean of all acquisitions. This
measurement is also affected by the nonlinearity inherent in the measurement
itself, as shown in fig.\ \ref{fig-mie}b.

\bibliography{refs}

\begin{thebibliography}{36}%
\makeatletter
\providecommand \@ifxundefined [1]{%
 \@ifx{#1\undefined}
}%
\providecommand \@ifnum [1]{%
 \ifnum #1\expandafter \@firstoftwo
 \else \expandafter \@secondoftwo
 \fi
}%
\providecommand \@ifx [1]{%
 \ifx #1\expandafter \@firstoftwo
 \else \expandafter \@secondoftwo
 \fi
}%
\providecommand \natexlab [1]{#1}%
\providecommand \enquote  [1]{``#1''}%
\providecommand \bibnamefont  [1]{#1}%
\providecommand \bibfnamefont [1]{#1}%
\providecommand \citenamefont [1]{#1}%
\providecommand \href@noop [0]{\@secondoftwo}%
\providecommand \href [0]{\begingroup \@sanitize@url \@href}%
\providecommand \@href[1]{\@@startlink{#1}\@@href}%
\providecommand \@@href[1]{\endgroup#1\@@endlink}%
\providecommand \@sanitize@url [0]{\catcode `\\12\catcode `\$12\catcode
  `\&12\catcode `\#12\catcode `\^12\catcode `\_12\catcode `\%12\relax}%
\providecommand \@@startlink[1]{}%
\providecommand \@@endlink[0]{}%
\providecommand \url  [0]{\begingroup\@sanitize@url \@url }%
\providecommand \@url [1]{\endgroup\@href {#1}{\urlprefix }}%
\providecommand \urlprefix  [0]{URL }%
\providecommand \Eprint [0]{\href }%
\providecommand \doibase [0]{http://dx.doi.org/}%
\providecommand \selectlanguage [0]{\@gobble}%
\providecommand \bibinfo  [0]{\@secondoftwo}%
\providecommand \bibfield  [0]{\@secondoftwo}%
\providecommand \translation [1]{[#1]}%
\providecommand \BibitemOpen [0]{}%
\providecommand \bibitemStop [0]{}%
\providecommand \bibitemNoStop [0]{.\EOS\space}%
\providecommand \EOS [0]{\spacefactor3000\relax}%
\providecommand \BibitemShut  [1]{\csname bibitem#1\endcsname}%
\let\auto@bib@innerbib\@empty
\bibitem [{\citenamefont {Murshed}\ \emph {et~al.}(2011)\citenamefont
  {Murshed}, \citenamefont {Nieto~de Castro}, \citenamefont {Louren{\c{c}}o},
  \citenamefont {Lopes},\ and\ \citenamefont {Santos}}]{murshed_review_2011}%
  \BibitemOpen
  \bibfield  {author} {\bibinfo {author} {\bibfnamefont {S.~M.~S.}\
  \bibnamefont {Murshed}}, \bibinfo {author} {\bibfnamefont {C.~A.}\
  \bibnamefont {Nieto~de Castro}}, \bibinfo {author} {\bibfnamefont {M.~J.~V.}\
  \bibnamefont {Louren{\c{c}}o}}, \bibinfo {author} {\bibfnamefont {M.~L.~M.}\
  \bibnamefont {Lopes}}, \ and\ \bibinfo {author} {\bibfnamefont {F.~J.~V.}\
  \bibnamefont {Santos}},\ }\href {\doibase 10.1016/j.rser.2011.02.016}
  {\bibfield  {journal} {\bibinfo  {journal} {Renew. Sust. Energ. Rev.}\
  }\textbf {\bibinfo {volume} {15}},\ \bibinfo {pages} {2342} (\bibinfo {year}
  {2011})}\BibitemShut {NoStop}%
\bibitem [{\citenamefont {Taylor}\ and\ \citenamefont
  {Phelan}(2009)}]{taylor_pool_2009}%
  \BibitemOpen
  \bibfield  {author} {\bibinfo {author} {\bibfnamefont {R.~A.}\ \bibnamefont
  {Taylor}}\ and\ \bibinfo {author} {\bibfnamefont {P.~E.}\ \bibnamefont
  {Phelan}},\ }\href {\doibase 10.1016/j.ijheatmasstransfer.2009.06.040}
  {\bibfield  {journal} {\bibinfo  {journal} {Int. J. Heat Mass Transfer}\
  }\textbf {\bibinfo {volume} {52}},\ \bibinfo {pages} {5339} (\bibinfo {year}
  {2009})}\BibitemShut {NoStop}%
\bibitem [{\citenamefont {Kim}\ \emph {et~al.}(2015)\citenamefont {Kim},
  \citenamefont {Yu}, \citenamefont {Jerng}, \citenamefont {Kim},\ and\
  \citenamefont {Ahn}}]{kim_review_2015}%
  \BibitemOpen
  \bibfield  {author} {\bibinfo {author} {\bibfnamefont {D.~E.}\ \bibnamefont
  {Kim}}, \bibinfo {author} {\bibfnamefont {D.~I.}\ \bibnamefont {Yu}},
  \bibinfo {author} {\bibfnamefont {D.~W.}\ \bibnamefont {Jerng}}, \bibinfo
  {author} {\bibfnamefont {M.~H.}\ \bibnamefont {Kim}}, \ and\ \bibinfo
  {author} {\bibfnamefont {H.~S.}\ \bibnamefont {Ahn}},\ }\href {\doibase
  10.1016/j.expthermflusci.2015.03.023} {\bibfield  {journal} {\bibinfo
  {journal} {Exp. Therm. Fluid Sci.}\ }\textbf {\bibinfo {volume} {66}},\
  \bibinfo {pages} {173} (\bibinfo {year} {2015})}\BibitemShut {NoStop}%
\bibitem [{\citenamefont {{\c{C}}engel}(2003)}]{cengel_pool_2003}%
  \BibitemOpen
  \bibfield  {author} {\bibinfo {author} {\bibfnamefont {Y.~A.}\ \bibnamefont
  {{\c{C}}engel}},\ }in\ \href@noop {} {{\selectlanguage {EN}\emph {\bibinfo
  {booktitle} {Heat {Transfer}: {A} {Practical} {Approach}}}}}\ (\bibinfo
  {publisher} {McGraw-Hill},\ \bibinfo {address} {Boston},\ \bibinfo {year}
  {2003})\ pp.\ \bibinfo {pages} {518--530}\BibitemShut {NoStop}%
\bibitem [{\citenamefont {Pioro}\ \emph
  {et~al.}(2004{\natexlab{a}})\citenamefont {Pioro}, \citenamefont {Rohsenow},\
  and\ \citenamefont {Doerffer}}]{pioro_nucleate_2004}%
  \BibitemOpen
  \bibfield  {author} {\bibinfo {author} {\bibfnamefont {I.~L.}\ \bibnamefont
  {Pioro}}, \bibinfo {author} {\bibfnamefont {W.}~\bibnamefont {Rohsenow}}, \
  and\ \bibinfo {author} {\bibfnamefont {S.~S.}\ \bibnamefont {Doerffer}},\
  }\href {\doibase 10.1016/j.ijheatmasstransfer.2004.06.019} {\bibfield
  {journal} {\bibinfo  {journal} {Int. J. Heat Mass Transfer}\ }\textbf
  {\bibinfo {volume} {47}},\ \bibinfo {pages} {5033} (\bibinfo {year}
  {2004}{\natexlab{a}})}\BibitemShut {NoStop}%
\bibitem [{\citenamefont {Pioro}\ \emph
  {et~al.}(2004{\natexlab{b}})\citenamefont {Pioro}, \citenamefont {Rohsenow},\
  and\ \citenamefont {Doerffer}}]{pioro_nucleate_2004-1}%
  \BibitemOpen
  \bibfield  {author} {\bibinfo {author} {\bibfnamefont {I.~L.}\ \bibnamefont
  {Pioro}}, \bibinfo {author} {\bibfnamefont {W.}~\bibnamefont {Rohsenow}}, \
  and\ \bibinfo {author} {\bibfnamefont {S.~S.}\ \bibnamefont {Doerffer}},\
  }\href {\doibase 10.1016/j.ijheatmasstransfer.2004.06.020} {\bibfield
  {journal} {\bibinfo  {journal} {Int. J. Heat Mass Transfer}\ }\textbf
  {\bibinfo {volume} {47}},\ \bibinfo {pages} {5045} (\bibinfo {year}
  {2004}{\natexlab{b}})}\BibitemShut {NoStop}%
\bibitem [{\citenamefont {Dong}\ \emph {et~al.}(2014)\citenamefont {Dong},
  \citenamefont {Quan},\ and\ \citenamefont {Cheng}}]{dong_experimental_2014}%
  \BibitemOpen
  \bibfield  {author} {\bibinfo {author} {\bibfnamefont {L.}~\bibnamefont
  {Dong}}, \bibinfo {author} {\bibfnamefont {X.}~\bibnamefont {Quan}}, \ and\
  \bibinfo {author} {\bibfnamefont {P.}~\bibnamefont {Cheng}},\ }\href
  {\doibase 10.1016/j.ijheatmasstransfer.2013.11.068} {\bibfield  {journal}
  {\bibinfo  {journal} {Int. J. Heat Mass Transfer}\ }\textbf {\bibinfo
  {volume} {71}},\ \bibinfo {pages} {189} (\bibinfo {year} {2014})}\BibitemShut
  {NoStop}%
\bibitem [{\citenamefont {Jakob}\ and\ \citenamefont
  {Linke}(1933)}]{jakob_warmeubergang_1933}%
  \BibitemOpen
  \bibfield  {author} {\bibinfo {author} {\bibfnamefont {M.}~\bibnamefont
  {Jakob}}\ and\ \bibinfo {author} {\bibfnamefont {W.}~\bibnamefont {Linke}},\
  }\href {\doibase 10.1007/BF02717048} {\bibfield  {journal} {\bibinfo
  {journal} {Forsch. Ing.-Wes.}\ }\textbf {\bibinfo {volume} {4}},\ \bibinfo
  {pages} {75} (\bibinfo {year} {1933})}\BibitemShut {NoStop}%
\bibitem [{\citenamefont {Jakob}\ and\ \citenamefont
  {Linke}(1935)}]{jakob_warmeubergang_1935}%
  \BibitemOpen
  \bibfield  {author} {\bibinfo {author} {\bibfnamefont {M.}~\bibnamefont
  {Jakob}}\ and\ \bibinfo {author} {\bibfnamefont {W.}~\bibnamefont {Linke}},\
  }\href@noop {} {\bibfield  {journal} {\bibinfo  {journal} {Phys. Z.}\
  }\textbf {\bibinfo {volume} {36}},\ \bibinfo {pages} {267} (\bibinfo {year}
  {1935})}\BibitemShut {NoStop}%
\bibitem [{\citenamefont {Leidenfrost}(1756)}]{leidenfrost_fixitate_1756}%
  \BibitemOpen
  \bibfield  {author} {\bibinfo {author} {\bibfnamefont {J.~G.}\ \bibnamefont
  {Leidenfrost}},\ }in\ \href@noop {} {{\selectlanguage {la}\emph {\bibinfo
  {booktitle} {De {Aqu{\ae}} {Communis} {Nonnullis} {Qualitatibus}
  {Tractatus}}}}}\ (\bibinfo {address} {Duisburgum ad Rhenum},\ \bibinfo {year}
  {1756})\ pp.\ \bibinfo {pages} {30--63}\BibitemShut {NoStop}%
\bibitem [{\citenamefont {Leidenfrost}(1966)}]{leidenfrost_fixation_1966}%
  \BibitemOpen
  \bibfield  {author} {\bibinfo {author} {\bibfnamefont {J.~G.}\ \bibnamefont
  {Leidenfrost}},\ }\href {\doibase 10.1016/0017-9310(66)90111-6} {\bibfield
  {journal} {\bibinfo  {journal} {Int. J. Heat Mass Transfer}\ }\textbf
  {\bibinfo {volume} {9}},\ \bibinfo {pages} {1153} (\bibinfo {year}
  {1966})}\BibitemShut {NoStop}%
\bibitem [{\citenamefont {Sher}\ \emph {et~al.}(2012)\citenamefont {Sher},
  \citenamefont {Harari}, \citenamefont {Reshef},\ and\ \citenamefont
  {Sher}}]{sher_film_2012}%
  \BibitemOpen
  \bibfield  {author} {\bibinfo {author} {\bibfnamefont {I.}~\bibnamefont
  {Sher}}, \bibinfo {author} {\bibfnamefont {R.}~\bibnamefont {Harari}},
  \bibinfo {author} {\bibfnamefont {R.}~\bibnamefont {Reshef}}, \ and\ \bibinfo
  {author} {\bibfnamefont {E.}~\bibnamefont {Sher}},\ }\href {\doibase
  10.1016/j.applthermaleng.2011.11.018} {\bibfield  {journal} {\bibinfo
  {journal} {Appl. Therm. Eng.}\ }\textbf {\bibinfo {volume} {36}},\ \bibinfo
  {pages} {219} (\bibinfo {year} {2012})}\BibitemShut {NoStop}%
\bibitem [{\citenamefont {Hou}\ \emph {et~al.}(2015)\citenamefont {Hou},
  \citenamefont {Yorulmaz}, \citenamefont {Verhart},\ and\ \citenamefont
  {Orrit}}]{hou_explosive_2015}%
  \BibitemOpen
  \bibfield  {author} {\bibinfo {author} {\bibfnamefont {L.}~\bibnamefont
  {Hou}}, \bibinfo {author} {\bibfnamefont {M.}~\bibnamefont {Yorulmaz}},
  \bibinfo {author} {\bibfnamefont {N.~R.}\ \bibnamefont {Verhart}}, \ and\
  \bibinfo {author} {\bibfnamefont {M.}~\bibnamefont {Orrit}},\ }\href
  {\doibase 10.1088/1367-2630/17/1/013050} {\bibfield  {journal} {\bibinfo
  {journal} {New J. Phys.}\ }\textbf {\bibinfo {volume} {17}},\ \bibinfo
  {pages} {013050} (\bibinfo {year} {2015})}\BibitemShut {NoStop}%
\bibitem [{\citenamefont {Baffou}\ \emph {et~al.}(2014)\citenamefont {Baffou},
  \citenamefont {Polleux}, \citenamefont {Rigneault},\ and\ \citenamefont
  {Monneret}}]{baffou_super-heating_2014}%
  \BibitemOpen
  \bibfield  {author} {\bibinfo {author} {\bibfnamefont {G.}~\bibnamefont
  {Baffou}}, \bibinfo {author} {\bibfnamefont {J.}~\bibnamefont {Polleux}},
  \bibinfo {author} {\bibfnamefont {H.}~\bibnamefont {Rigneault}}, \ and\
  \bibinfo {author} {\bibfnamefont {S.}~\bibnamefont {Monneret}},\ }\href
  {\doibase 10.1021/jp411519k} {\bibfield  {journal} {\bibinfo  {journal} {J.
  Phys. Chem. C}\ }\textbf {\bibinfo {volume} {118}},\ \bibinfo {pages} {4890}
  (\bibinfo {year} {2014})}\BibitemShut {NoStop}%
\bibitem [{\citenamefont {Siems}\ \emph {et~al.}(2011)\citenamefont {Siems},
  \citenamefont {Weber}, \citenamefont {Boneberg},\ and\ \citenamefont
  {Plech}}]{siems_thermodynamics_2011}%
  \BibitemOpen
  \bibfield  {author} {\bibinfo {author} {\bibfnamefont {A.}~\bibnamefont
  {Siems}}, \bibinfo {author} {\bibfnamefont {S.~A.~L.}\ \bibnamefont {Weber}},
  \bibinfo {author} {\bibfnamefont {J.}~\bibnamefont {Boneberg}}, \ and\
  \bibinfo {author} {\bibfnamefont {A.}~\bibnamefont {Plech}},\ }\href
  {\doibase 10.1088/1367-2630/13/4/043018} {\bibfield  {journal} {\bibinfo
  {journal} {New J. Phys.}\ }\textbf {\bibinfo {volume} {13}},\ \bibinfo
  {pages} {043018} (\bibinfo {year} {2011})}\BibitemShut {NoStop}%
\bibitem [{\citenamefont {Boulais}\ \emph {et~al.}(2012)\citenamefont
  {Boulais}, \citenamefont {Lachaine},\ and\ \citenamefont
  {Meunier}}]{boulais_plasma_2012}%
  \BibitemOpen
  \bibfield  {author} {\bibinfo {author} {\bibfnamefont {{\'E}.}~\bibnamefont
  {Boulais}}, \bibinfo {author} {\bibfnamefont {R.}~\bibnamefont {Lachaine}}, \
  and\ \bibinfo {author} {\bibfnamefont {M.}~\bibnamefont {Meunier}},\ }\href
  {\doibase 10.1021/nl302200w} {\bibfield  {journal} {\bibinfo  {journal} {Nano
  Lett.}\ }\textbf {\bibinfo {volume} {12}},\ \bibinfo {pages} {4763} (\bibinfo
  {year} {2012})}\BibitemShut {NoStop}%
\bibitem [{\citenamefont {Hashimoto}\ \emph {et~al.}(2012)\citenamefont
  {Hashimoto}, \citenamefont {Werner},\ and\ \citenamefont
  {Uwada}}]{hashimoto_studies_2012}%
  \BibitemOpen
  \bibfield  {author} {\bibinfo {author} {\bibfnamefont {S.}~\bibnamefont
  {Hashimoto}}, \bibinfo {author} {\bibfnamefont {D.}~\bibnamefont {Werner}}, \
  and\ \bibinfo {author} {\bibfnamefont {T.}~\bibnamefont {Uwada}},\ }\href
  {\doibase 10.1016/j.jphotochemrev.2012.01.001} {\bibfield  {journal}
  {\bibinfo  {journal} {J. Photochem. Photobiol. C}\ }\textbf {\bibinfo
  {volume} {13}},\ \bibinfo {pages} {28} (\bibinfo {year} {2012})}\BibitemShut
  {NoStop}%
\bibitem [{\citenamefont {Lombard}\ \emph {et~al.}(2014)\citenamefont
  {Lombard}, \citenamefont {Biben},\ and\ \citenamefont
  {Merabia}}]{lombard_kinetics_2014}%
  \BibitemOpen
  \bibfield  {author} {\bibinfo {author} {\bibfnamefont {J.}~\bibnamefont
  {Lombard}}, \bibinfo {author} {\bibfnamefont {T.}~\bibnamefont {Biben}}, \
  and\ \bibinfo {author} {\bibfnamefont {S.}~\bibnamefont {Merabia}},\ }\href
  {\doibase 10.1103/PhysRevLett.112.105701} {\bibfield  {journal} {\bibinfo
  {journal} {Phys. Rev. Lett.}\ }\textbf {\bibinfo {volume} {112}},\ \bibinfo
  {pages} {105701} (\bibinfo {year} {2014})}\BibitemShut {NoStop}%
\bibitem [{\citenamefont {Lombard}\ \emph {et~al.}(2015)\citenamefont
  {Lombard}, \citenamefont {Biben},\ and\ \citenamefont
  {Merabia}}]{lombard_nanobubbles_2015}%
  \BibitemOpen
  \bibfield  {author} {\bibinfo {author} {\bibfnamefont {J.}~\bibnamefont
  {Lombard}}, \bibinfo {author} {\bibfnamefont {T.}~\bibnamefont {Biben}}, \
  and\ \bibinfo {author} {\bibfnamefont {S.}~\bibnamefont {Merabia}},\ }\href
  {\doibase 10.1103/PhysRevE.91.043007} {\bibfield  {journal} {\bibinfo
  {journal} {Phys. Rev. E}\ }\textbf {\bibinfo {volume} {91}},\ \bibinfo
  {pages} {043007} (\bibinfo {year} {2015})}\BibitemShut {NoStop}%
\bibitem [{\citenamefont {Setoura}\ \emph {et~al.}(2017)\citenamefont
  {Setoura}, \citenamefont {Ito},\ and\ \citenamefont
  {Miyasaka}}]{setoura_stationary_2017}%
  \BibitemOpen
  \bibfield  {author} {\bibinfo {author} {\bibfnamefont {K.}~\bibnamefont
  {Setoura}}, \bibinfo {author} {\bibfnamefont {S.}~\bibnamefont {Ito}}, \ and\
  \bibinfo {author} {\bibfnamefont {H.}~\bibnamefont {Miyasaka}},\ }\href
  {\doibase 10.1039/C6NR07990C} {\bibfield  {journal} {\bibinfo  {journal}
  {Nanoscale}\ }\textbf {\bibinfo {volume} {9}},\ \bibinfo {pages} {719}
  (\bibinfo {year} {2017})}\BibitemShut {NoStop}%
\bibitem [{\citenamefont {Li}\ \emph {et~al.}(2017)\citenamefont {Li},
  \citenamefont {Gonzalez-Avila}, \citenamefont {Nguyen},\ and\ \citenamefont
  {Ohl}}]{li_oscillate_2017}%
  \BibitemOpen
  \bibfield  {author} {\bibinfo {author} {\bibfnamefont {F.}~\bibnamefont
  {Li}}, \bibinfo {author} {\bibfnamefont {S.~R.}\ \bibnamefont
  {Gonzalez-Avila}}, \bibinfo {author} {\bibfnamefont {D.~M.}\ \bibnamefont
  {Nguyen}}, \ and\ \bibinfo {author} {\bibfnamefont {C.-D.}\ \bibnamefont
  {Ohl}},\ }\href {\doibase 10.1103/PhysRevFluids.2.014007} {\bibfield
  {journal} {\bibinfo  {journal} {Phys. Rev. Fluids}\ }\textbf {\bibinfo
  {volume} {2}},\ \bibinfo {pages} {014007} (\bibinfo {year}
  {2017})}\BibitemShut {NoStop}%
\bibitem [{\citenamefont {Gaiduk}\ \emph {et~al.}(2010)\citenamefont {Gaiduk},
  \citenamefont {V.~Ruijgrok}, \citenamefont {Yorulmaz},\ and\ \citenamefont
  {Orrit}}]{gaiduk_detection_2010}%
  \BibitemOpen
  \bibfield  {author} {\bibinfo {author} {\bibfnamefont {A.}~\bibnamefont
  {Gaiduk}}, \bibinfo {author} {\bibfnamefont {P.}~\bibnamefont {V.~Ruijgrok}},
  \bibinfo {author} {\bibfnamefont {M.}~\bibnamefont {Yorulmaz}}, \ and\
  \bibinfo {author} {\bibfnamefont {M.}~\bibnamefont {Orrit}},\ }\href
  {\doibase 10.1039/C0SC00210K} {\bibfield  {journal} {\bibinfo  {journal}
  {Chem. Sci.}\ }\textbf {\bibinfo {volume} {1}},\ \bibinfo {pages} {343}
  (\bibinfo {year} {2010})}\BibitemShut {NoStop}%
\bibitem [{\citenamefont {Jollans}\ \emph {et~al.}(2019)\citenamefont
  {Jollans}, \citenamefont {Baaske},\ and\ \citenamefont
  {Orrit}}]{jollans_nonfluorescent_2019}%
  \BibitemOpen
  \bibfield  {author} {\bibinfo {author} {\bibfnamefont {T.}~\bibnamefont
  {Jollans}}, \bibinfo {author} {\bibfnamefont {M.~D.}\ \bibnamefont {Baaske}},
  \ and\ \bibinfo {author} {\bibfnamefont {M.}~\bibnamefont {Orrit}},\ }\href
  {\doibase 10.1021/acs.jpcc.9b00843} {\bibfield  {journal} {\bibinfo
  {journal} {J. Phys. Chem. C}\ }\textbf {\bibinfo {volume} {123}},\ \bibinfo
  {pages} {14107} (\bibinfo {year} {2019})}\BibitemShut {NoStop}%
\bibitem [{\citenamefont {Pe{\~n}a}\ and\ \citenamefont
  {Pal}(2009)}]{pena_scattering_2009}%
  \BibitemOpen
  \bibfield  {author} {\bibinfo {author} {\bibfnamefont {O.}~\bibnamefont
  {Pe{\~n}a}}\ and\ \bibinfo {author} {\bibfnamefont {U.}~\bibnamefont {Pal}},\
  }\href {\doibase 10.1016/j.cpc.2009.07.010} {\bibfield  {journal} {\bibinfo
  {journal} {Comput. Phys. Comm.}\ }\textbf {\bibinfo {volume} {180}},\
  \bibinfo {pages} {2348} (\bibinfo {year} {2009})}\BibitemShut {NoStop}%
\bibitem [{\citenamefont {Setoura}\ \emph {et~al.}(2014)\citenamefont
  {Setoura}, \citenamefont {Okada},\ and\ \citenamefont
  {Hashimoto}}]{setoura_cw-laser-induced_2014}%
  \BibitemOpen
  \bibfield  {author} {\bibinfo {author} {\bibfnamefont {K.}~\bibnamefont
  {Setoura}}, \bibinfo {author} {\bibfnamefont {Y.}~\bibnamefont {Okada}}, \
  and\ \bibinfo {author} {\bibfnamefont {S.}~\bibnamefont {Hashimoto}},\ }\href
  {\doibase 10.1039/C4CP03733B} {\bibfield  {journal} {\bibinfo  {journal}
  {Phys. Chem. Chem. Phys.}\ }\textbf {\bibinfo {volume} {16}},\ \bibinfo
  {pages} {26938} (\bibinfo {year} {2014})}\BibitemShut {NoStop}%
\bibitem [{\citenamefont {Hashimoto}\ \emph {et~al.}(2009)\citenamefont
  {Hashimoto}, \citenamefont {Uwada}, \citenamefont {Hagiri}, \citenamefont
  {Takai},\ and\ \citenamefont {Ueki}}]{hashimoto_gold_2009}%
  \BibitemOpen
  \bibfield  {author} {\bibinfo {author} {\bibfnamefont {S.}~\bibnamefont
  {Hashimoto}}, \bibinfo {author} {\bibfnamefont {T.}~\bibnamefont {Uwada}},
  \bibinfo {author} {\bibfnamefont {M.}~\bibnamefont {Hagiri}}, \bibinfo
  {author} {\bibfnamefont {H.}~\bibnamefont {Takai}}, \ and\ \bibinfo {author}
  {\bibfnamefont {T.}~\bibnamefont {Ueki}},\ }\href {\doibase
  10.1021/jp905291h} {\bibfield  {journal} {\bibinfo  {journal} {J. Phys. Chem.
  C}\ }\textbf {\bibinfo {volume} {113}},\ \bibinfo {pages} {20640} (\bibinfo
  {year} {2009})}\BibitemShut {NoStop}%
\bibitem [{\citenamefont {Lauterborn}\ and\ \citenamefont
  {Kurz}(2010)}]{lauterborn_physics_2010}%
  \BibitemOpen
  \bibfield  {author} {\bibinfo {author} {\bibfnamefont {W.}~\bibnamefont
  {Lauterborn}}\ and\ \bibinfo {author} {\bibfnamefont {T.}~\bibnamefont
  {Kurz}},\ }\href {\doibase 10.1088/0034-4885/73/10/106501} {\bibfield
  {journal} {\bibinfo  {journal} {Rep. Prog. Phys.}\ }\textbf {\bibinfo
  {volume} {73}},\ \bibinfo {pages} {106501} (\bibinfo {year}
  {2010})}\BibitemShut {NoStop}%
\bibitem [{noa(2018)}]{noauthor_pentane_2018}%
  \BibitemOpen
  \href {https://www.engineeringtoolbox.com/pentane-properties-d_2048.html}
  {\enquote {\bibinfo {title} {Pentane -- {Thermophysical} {Properties}},}\
  }\bibinfo {howpublished} {in \textit{Engineering {ToolBox}}} (\bibinfo {year}
  {2018}),\ \bibinfo {note} {[online]
  \url{https://www.engineeringtoolbox.com/pentane-properties-d_2048.html}.
  Accessed 2019-01-28}\BibitemShut {NoStop}%
\bibitem [{\citenamefont {Vesovic}()}]{vesovic_pentane_2011}%
  \BibitemOpen
  \bibfield  {author} {\bibinfo {author} {\bibfnamefont {V.}~\bibnamefont
  {Vesovic}},\ }\href {\doibase 10.1615/AtoZ.p.pentane} {\enquote {\bibinfo
  {title} {{PENTANE}},}\ }\bibinfo {howpublished} {in \emph{Thermopedia}},\
  \bibinfo {note} {[online] \url{http://www.thermopedia.com/content/1016/}.
  Accessed 2019-02-13.}\BibitemShut {Stop}%
\bibitem [{\citenamefont {Minnaert}(1933)}]{minnaert_musical_1933}%
  \BibitemOpen
  \bibfield  {author} {\bibinfo {author} {\bibfnamefont {M.}~\bibnamefont
  {Minnaert}},\ }\href {\doibase 10.1080/14786443309462277} {\bibfield
  {journal} {\bibinfo  {journal} {London Edinburgh Dublin Philos. Mag. J.
  Sci.}\ }\textbf {\bibinfo {volume} {16}},\ \bibinfo {pages} {235} (\bibinfo
  {year} {1933})}\BibitemShut {NoStop}%
\bibitem [{Note1()}]{Note1}%
  \BibitemOpen
  \bibinfo {note} {See also Sec.\ 11 of the Supplementary Information of Ref.
  \cite {hou_explosive_2015}.}\BibitemShut {Stop}%
\bibitem [{\citenamefont {MacDonald}\ \emph {et~al.}(2003)\citenamefont
  {MacDonald}, \citenamefont {Fedotov}, \citenamefont {Pochon}, \citenamefont
  {Soares}, \citenamefont {Zheludev}, \citenamefont {Guignard}, \citenamefont
  {Mihaescu},\ and\ \citenamefont {Besnard}}]{macdonald_oscillating_2003}%
  \BibitemOpen
  \bibfield  {author} {\bibinfo {author} {\bibfnamefont {K.~F.}\ \bibnamefont
  {MacDonald}}, \bibinfo {author} {\bibfnamefont {V.~A.}\ \bibnamefont
  {Fedotov}}, \bibinfo {author} {\bibfnamefont {S.}~\bibnamefont {Pochon}},
  \bibinfo {author} {\bibfnamefont {B.~F.}\ \bibnamefont {Soares}}, \bibinfo
  {author} {\bibfnamefont {N.~I.}\ \bibnamefont {Zheludev}}, \bibinfo {author}
  {\bibfnamefont {C.}~\bibnamefont {Guignard}}, \bibinfo {author}
  {\bibfnamefont {A.}~\bibnamefont {Mihaescu}}, \ and\ \bibinfo {author}
  {\bibfnamefont {P.}~\bibnamefont {Besnard}},\ }\href {\doibase
  10.1103/PhysRevE.68.027301} {\bibfield  {journal} {\bibinfo  {journal} {Phys.
  Rev. E}\ }\textbf {\bibinfo {volume} {68}},\ \bibinfo {pages} {027301}
  (\bibinfo {year} {2003})}\BibitemShut {NoStop}%
\bibitem [{\citenamefont {Zharov}\ \emph {et~al.}(2005)\citenamefont {Zharov},
  \citenamefont {Letfullin},\ and\ \citenamefont
  {Galitovskaya}}]{zharov_microbubbles-overlapping_2005}%
  \BibitemOpen
  \bibfield  {author} {\bibinfo {author} {\bibfnamefont {V.~P.}\ \bibnamefont
  {Zharov}}, \bibinfo {author} {\bibfnamefont {R.~R.}\ \bibnamefont
  {Letfullin}}, \ and\ \bibinfo {author} {\bibfnamefont {E.~N.}\ \bibnamefont
  {Galitovskaya}},\ }\href {\doibase 10.1088/0022-3727/38/15/007} {\bibfield
  {journal} {\bibinfo  {journal} {J. Phys. D}\ }\textbf {\bibinfo {volume}
  {38}},\ \bibinfo {pages} {2571} (\bibinfo {year} {2005})}\BibitemShut
  {NoStop}%
\bibitem [{\citenamefont {Huang}\ \emph {et~al.}(2008)\citenamefont {Huang},
  \citenamefont {Jain}, \citenamefont {El-Sayed},\ and\ \citenamefont
  {El-Sayed}}]{huang_plasmonic_2008}%
  \BibitemOpen
  \bibfield  {author} {\bibinfo {author} {\bibfnamefont {X.}~\bibnamefont
  {Huang}}, \bibinfo {author} {\bibfnamefont {P.~K.}\ \bibnamefont {Jain}},
  \bibinfo {author} {\bibfnamefont {I.~H.}\ \bibnamefont {El-Sayed}}, \ and\
  \bibinfo {author} {\bibfnamefont {M.~A.}\ \bibnamefont {El-Sayed}},\ }\href
  {\doibase 10.1007/s10103-007-0470-x} {\bibfield  {journal} {\bibinfo
  {journal} {Lasers Med. Sci.}\ }\textbf {\bibinfo {volume} {23}},\ \bibinfo
  {pages} {217} (\bibinfo {year} {2008})}\BibitemShut {NoStop}%
\bibitem [{\citenamefont {Lapotko}(2011)}]{lapotko_plasmonic_2011}%
  \BibitemOpen
  \bibfield  {author} {\bibinfo {author} {\bibfnamefont {D.}~\bibnamefont
  {Lapotko}},\ }\href {\doibase 10.3390/cancers3010802} {\bibfield  {journal}
  {\bibinfo  {journal} {Cancers}\ }\textbf {\bibinfo {volume} {3}},\ \bibinfo
  {pages} {802} (\bibinfo {year} {2011})}\BibitemShut {NoStop}%
\bibitem [{\citenamefont {Funke}\ \emph {et~al.}(1997)\citenamefont {Funke},
  \citenamefont {Argo}, \citenamefont {Falconer},\ and\ \citenamefont
  {Noble}}]{funke_separations_1997}%
  \BibitemOpen
  \bibfield  {author} {\bibinfo {author} {\bibfnamefont {H.~H.}\ \bibnamefont
  {Funke}}, \bibinfo {author} {\bibfnamefont {A.~M.}\ \bibnamefont {Argo}},
  \bibinfo {author} {\bibfnamefont {J.~L.}\ \bibnamefont {Falconer}}, \ and\
  \bibinfo {author} {\bibfnamefont {R.~D.}\ \bibnamefont {Noble}},\ }\href
  {\doibase 10.1021/ie960472f} {\bibfield  {journal} {\bibinfo  {journal} {Ind.
  Eng. Chem. Res.}\ }\textbf {\bibinfo {volume} {36}},\ \bibinfo {pages} {137}
  (\bibinfo {year} {1997})}\BibitemShut {NoStop}%
\end{thebibliography}%

\end{document}